\documentclass[%
reprint,
 amsmath,amssymb,
 aps,
]{revtex4-2}
\pdfoutput=1
 
\usepackage{graphicx}
\usepackage{dcolumn}
\usepackage{bm}

\begin{document}

\preprint{APS/123-QED}

\title{Entry of Microparticles into Giant Lipid Vesicles by Optical Tweezers}

\author{Florent Fessler}
 \email{florent.fessler@ics-cnrs.unistra.fr}
 \author{Vaibhav Sharma}
 \author{Pierre Muller}
\author{Antonio Stocco}%
 \email{stocco@unistra.fr}
\affiliation{%
 Institut Charles Sadron, CNRS UPR 22, 23 rue du Loess, 67200 Strasbourg, France}%


%
%


\date{\today}

\begin{abstract}
Entry of micro- or nano-sized objects into cells or vesicles made of lipid membranes occur in many processes such as entry of viruses in host cells, microplastics pollution, drug delivery or biomedical imaging. 
Here, we investigated the microparticle crossing of lipid membranes in giant unilamellar vesicles in the absence of strong binding interactions (e.g. streptavidin-biotin binding). In these conditions, we observed that organic and inorganic particles can always penetrate inside the vesicles provided that an external picoNewton force is applied and for relatively low membrane tensions. In the limit of a vanishing adhesion, we pointed out the role of the membrane area reservoir and show that a force minimum exists when the particle size is comparable to the bendocapillary length.

\end{abstract}



\maketitle

The interactions between micro- or nano-sized bodies and lipid membranes govern many mechanisms taking place in many fields such as viral infection, drug delivery or biomedical imaging \cite{Mercer2010,Negi2022,Kukulski2012}. The relation between the physical properties of these soft fluctuating membranes and their shape transitions upon interaction with a particle constitutes an active domain of investigations \cite{Frey2021, Jaggers2018, Koch2020, Kim2020} and has been extensively studied theoretically \cite{Deserno2004,Bahrami2016,Lipo1,Lipowsky_1998,Lipowsky2012,Benoit2007,Jaime2021,Xiao2022}.

Generally, a membrane can be deformed upon contact with a particle as a result of an adhesion energy $E_{w}$ driving the wrapping of the object competing against the energies resisting the deformation associated to the tension $E_{\sigma}$ and the bending $E_{b}$ of the membrane. These energies can be expressed as a function of the system properties, namely the membrane tension $\sigma$, the membrane bending rigidity $\kappa_b$ and membrane-particle adhesive energy per unit area $w$, which is negative for the spontaneous particle wrapping by the membrane. Here we consider a force driven particle wrapping as a consequence of the nucleation and formation of a membrane neck or tube, which may occur even for vanishing or energetically unfavorable particle-membrane adhesion ($w\geq$ 0)  \cite{Deserno2004,Koster2005,Prost2002,Powers2002,Frolov2003}.

Few experimental investigations were able to address the wrapping of particles by giant unilamellar vesicles (GUVs) \cite{Spanke2020,Spanke2021, Dietrich1997}, and only recently insights into the physics of particle wrapping have been reported in some specific experimental regimes. In the limit of negligible membrane tensions and by triggering the attraction between particles and membranes using polymer depletants, Spanke $\it{et}$ $\it{al.}$ have shown that spontaneous or activated particle wrapping by a lipid vesicle can occur \cite{Spanke2020,Spanke2021}. Experiments were described by considering lengthscales such as the bendocapillary length $\lambda_\sigma=\sqrt{\kappa_b/\sigma}$ capturing the competition between membrane bending and tension, and the adhesion length $\lambda_w=\sqrt{2\kappa_b/w}$ describing  the competition between  bending and adhesion. Their experiments were carried out in a regime governed by the membrane bending, where the particle radius $R_P < \lambda_\sigma$. Note that in most experimental studies, the tension considered only accounts for a mechanical tension. However, it was predicted that membranes with a spontaneous curvature will show a spontaneous tension $\tilde{\sigma}=2\kappa m^2$ contributing and adding up to the mechanical tension $\Sigma$, leading to $\sigma= \Sigma+\tilde{\sigma}$ \cite{Lipowsky2012,Jaime2021}.

The existence of an unexpectedly robust \textit{activated wrapping} regime \cite{Spanke2020} for vanishing membrane tensions 
was attributed to the membrane finite mean curvature initially bulging towards the particle \cite{Bahrami2016,Lipo1}, while other energy barriers may exist preventing spontaneous particle wrapping by the membrane \cite{Deserno2004, Deserno2004_2}. In a regime governed by the membrane tension, $R_P > \lambda_\sigma $ with $ \sigma \approx 10^{-6}\  \text{N.m}^{-1}$ (negligible bending), wrapping of particles by pushing them across a free-standing membrane using optical forces \cite{DolsPerez2019} was also observed.

Considering vesicles possessing low membrane tensions with $\sigma \approx 10^{-7}-10^{-8}$ N.m$^{-1}$  and a  typical bending rigidity $\kappa_b \approx 1\times10^{-19} $ J , one finds that the bendocapillary length lays in the range 1 $\mu$m $< \lambda_\sigma < \ $3 $\mu$m.  Hence, membrane deformations induced by particles with radii $R_P$ of the order of a micron lay in a crossover regime, where both tension and bending may significantly contribute to the total energy and the criterion $R_P > \lambda_w$ might not be sufficient to trigger the wrapping.

\begin{figure}[t]
\includegraphics[width=0.95\linewidth]{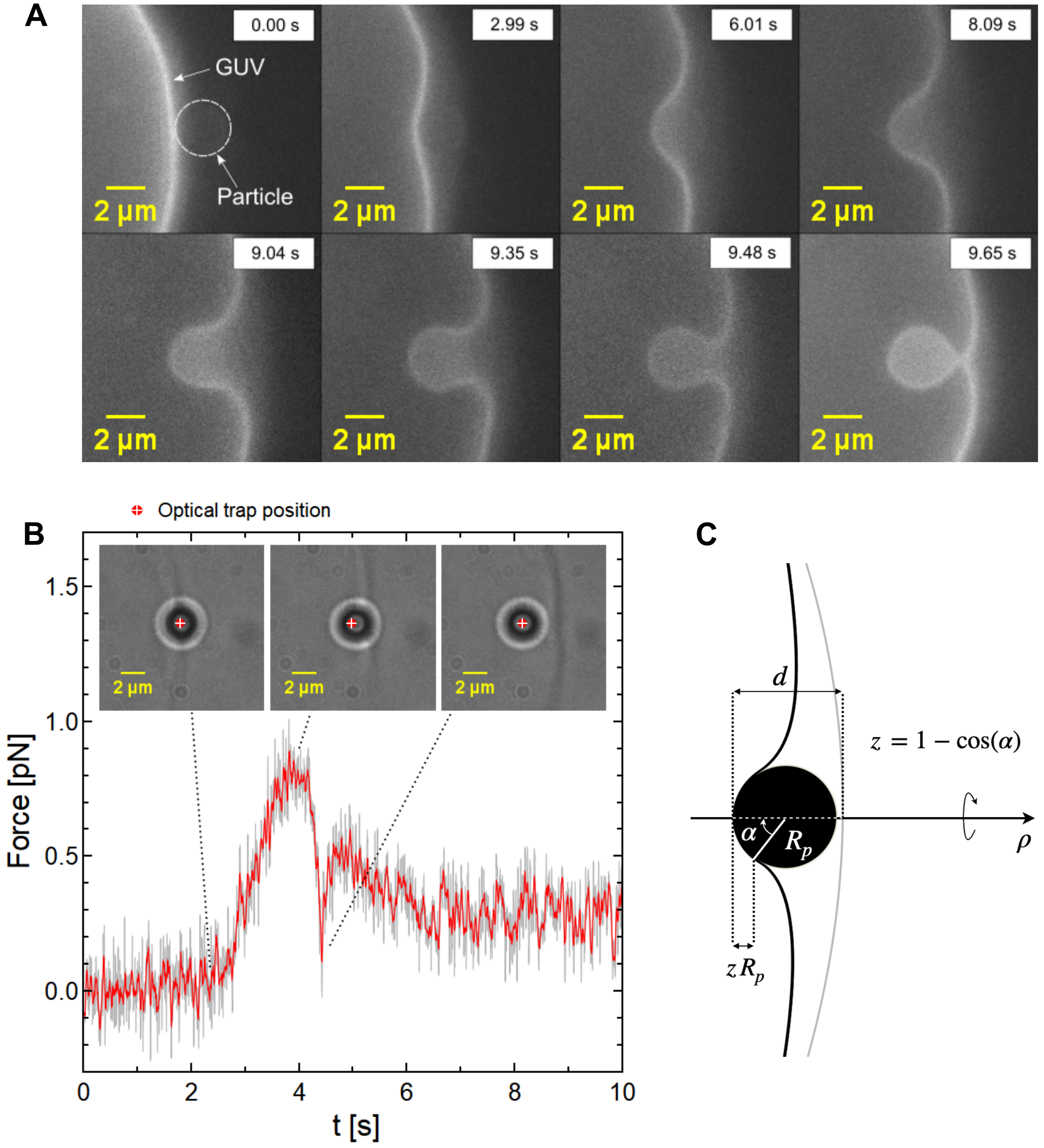}
\caption{ (\textit{A}) Fluorescence microscopy snapshots at key moments (time lag between each snapshot is not constant) of an optically trapped $R_P=1.15 \ \mu \text{m}$ Silica particle penetrating inside a GUV at $v_{rel} = 1.9 \pm 0.02 \ \mu \text{m.s}^{-1}$ (\textit{B}) A typical force profile with associated bright field microscopy snapshots at key moments of the penetration process. The grey curve is the raw signal and the red one is a sliding average over 20 points. (\textit{C}) Sketch defining the penetration degree $z$, contact angle $\alpha$ and penetration depth $d$.}
\label{fig1}
\end{figure}

In this Letter, we present experimental investigations combining force measurements with fluorescence as well as bright field optical imaging on the driven microparticle wrapping by giant unilamellar vesicles in a crossover regime where the particle size is comparable to the bendocapillary length, $R_P \approx \lambda_\sigma$. Our experiments allow to resolve the deformations of the membrane around the particle and measure the force during the penetration of the particle inside a GUV. We investigate the effects of membrane bending and tension contributions by using different particle sizes, and the effect of particle adhesion by using particles made of different materials. In order not to alter membrane properties such as  bending rigidity or spontaneous curvature, no additional depletants, salts or chemicals are added. The interactions between the particle and the membrane are therefore weak, reversible and non specific ($\left|w\right| < 10^{-7}\  \text{N.m}^{-1}$) \cite{Gruhn2007,MOY1999}.
1-palmitoyl-2-oleyl-sn-glycero-3-phosphocholine (POPC) GUVs with radii $R_v>10 \ \mu$m containing 1$\%$ Nitrobenzoxadiazole (NBD) head-labelled lipids as a fluorescent marker and commercial particles of different composititons (Silica, Melamine Formadehdyde and Polystyrene) and sizes in the micron range were used. The GUVs were formed using a gel-assisted formation method \cite{Weinberger2013} in 0.15 Osm/kg sucrose solution and sedimented in 0.15 Osm/kg glucose solution. Low tension vesicles were obtained by letting the sample evaporate for 1 hour before performing experiments thereby creating an osmotic imbalance between the inner and outer compartment of the GUVs leading to deflated low tension vesicles.

In Fig \ref{fig1}A we show typical dynamics observed in our experiments in fluorescence microscopy (App. \ref{microscopy}). We are able to image the lipid membrane deformations due to the penetration of the particle driven by optical tweezers and observe the shape transitions occurring until the complete particle wrapping by the GUV membrane (see Supplementary Video S1). The non-fluorescent Silica particle is optically trapped and the sample stage is moved with controlled speed (see App. \ref{penetrationspeed}) to approach the GUV and force the penetration of the particle. Note that if the optical trap is turned off before $t= 9$ s in Fig \ref{fig1}A, the Silica particle is expelled and the membrane recovers its initial shape. Therefore one can refer to this first reversible stage of the membrane deformation ($t< 9$ s ) as \textit{elastic}.
On the other hand, once the particle contact angle $\alpha$ (see Fig \ref{fig1}C) reaches $\alpha_c \approx \pi/2$, the membrane neck forms, and it is impossible to take the particle back out and unwrap it using the same optical force in the opposite direction. Hence, this second stage of the process can be considered as irreversible (App. \ref{irreversibility}). Time stamps on the snapshots in \ref{fig1}A highlight the difference in membrane shape transition timescales between the first step of the slow reversible deformation (elastic-like, first row of snapshots) whose dynamics is governed by the relative velocity between the two objects and the neck formation step whose dynamics seem to be dictated by the membrane dynamics as soon as the critical contact angle $\alpha_c$ is reached (second row of snapshots). 
This instability is analogous to first order shape transitions reported in tube pulling assays under some conditions \cite{Koster2005,Prost2002}.

Calibration of the optical trap allows to record the force exerted by the membrane on the particle upon penetration (App. \ref{calibration}). Fig. \ref{fig1}B shows a force profile associated to the forced penetration performed at constant speed. The force is plotted as a function of the time of the experiment, which can  be converted into a length (proportional to the penetration depth $d$). The fluctuations of the force throughout the experiment account for thermal fluctuations of the bath (random force). The profiles can be decomposed in distinct steps. First, the force oscillates around zero when the particle approaches the GUV. Indeed, the Stokes friction before contact $F_v=6\pi \eta R_P v_{rel} \approx$  0.01 pN can not be resolved and is therefore negligible in our experiments \cite{DolsPerez2019} (App. \ref{penetrationspeed}). Around $t=$ 3 s, the particle touches the GUV  and the force grows linearly in time reaching a maximum value $F_{M}$ corresponding to the end of the $\it{elastic}$ membrane deformation regime. The sharp drop of the force after the maximum corresponds to the formation of the neck and complete wrapping of the particle. If the optical trap is released when the force drops down to zero, the particle remains stably wrapped by the membrane. In our experiments, continuing the particle penetration inside the GUV leads to the formation of a membrane tube with associated plateau force $F_{tube}=f_{in}=2\pi\sqrt{2\kappa\sigma}+4\pi \kappa m$ (from $t \approx 6$ s to the end), where $f_{in}$ stands for the force needed to pull an inward tube \cite{Dasgupta2018}. The force rebound measured after the sharp force drop  associated to the wrapping of the particle  in Fig \ref{fig1}B accounts for the force barrier to go from the stable engulfed state with an ideal neck where membrane and particle are in contact \cite{Spanke2020} to a state where the fully wrapped particle is connected to the GUV with a tube. This is reminiscent of the situation where a point force is applied to a GUV to form outward tubes in which a force overshoot is measured just before the catenoid-like pulled membrane segment collapses into a tube \cite{Koster2005,Powers2002,Ashok2014,Roopa2003}. The geometry is however more complex here and the shape of the overshoot before the plateau do not suggest a first order shape transition.

\begin{figure}[t]
\includegraphics[width=1\linewidth]{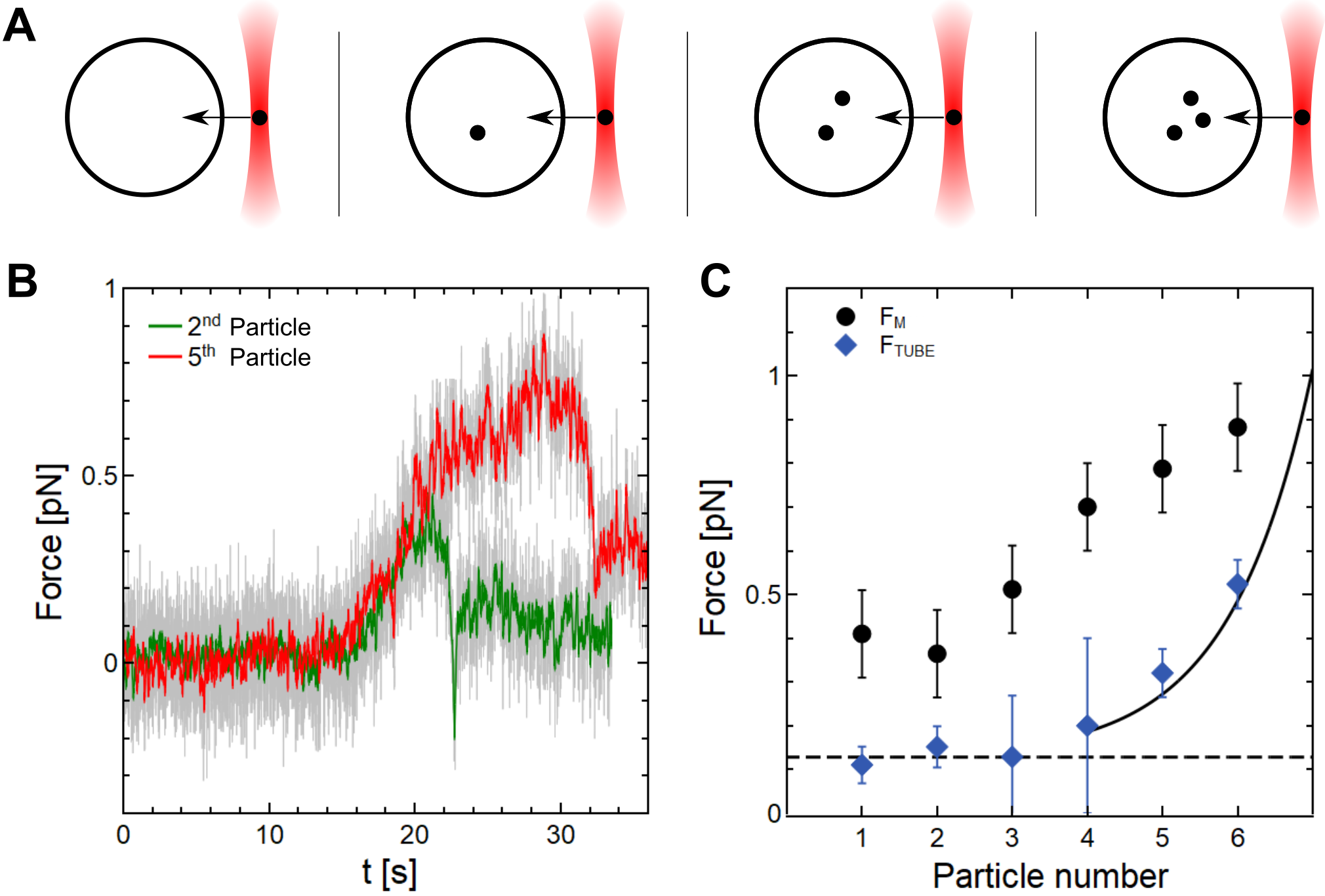}
\caption{\textit{(A)} Schematic representation of the experiment consisting in successively penetrating $R_P= 1.15 \ \mu$m Silica particles in a same GUV. \textit{(B)} Representative force profiles upon penetration for the second and fifth penetrated particles in a same GUV. \textit{(C)} Maximmum force $F_{M}$ and plateau force $F_{tube}$ measured for each successive particle.  The dashed black line is  $f_{in}$  for constant $\Sigma= 10^{-8}$ N.m$^{-1}$ and $m=2 \times 10^5$ m$^{-1}$ while the solid black curve stands for $f_{in}$ with $\Sigma$ following Eq. (\ref{eqnlnsigma}) for $\Sigma_0=10^{-9}$ N.m$^{-1}$ (see the text). 
 }
\label{fig2}
\end{figure}

The maximum force provided by optical trapping during penetration $F_{M}$ contains information on the energy barrier that has to be overcome to wrap a particle in absence of strong binding \cite{Koltover1999} and spontaneous wrapping. We start analyzing how $F_{M}$ relates to $F_{tube}$ and, in turn, to the membrane properties. To do so, we perform an experiment depicted in \ref{fig2}A which consists in successively penetrating $R_P=1.15\ \mu$m particles in the same GUV, thereby conserving all the properties of the system except the membrane area that is consumed by each particle remaining stably wrapped after penetration.  
Results in Fig \ref{fig2}B show two representative force profiles for this experiment and shows the correlated raise of both $F_{M}$ and $F_{tube}$ as well as an increase of the time $\tau_p$ between contact and neck closure. In this experiment the optical spring constant was fixed, and we were able to penetrate up to 6 particles but the 7th did not enter with the optical forces reachable at this fixed optical power.  As explained in the following, these differences reveal the impact of the membrane area reservoir and membrane tension on the energy barrier for wrapping. We postulate that the plateau force $F_{tube}$ depends only on the properties of the membrane and is independent on the particle interaction with the membrane. Thus, upon particle penetration the vesicle surface-to-volume ratio changes, which in turn affects the membrane tension. For the first 3 penetrating particles $F_{tube}$ does not vary, which could seem surprising since significant membrane area is consumed.
In fact, for spherical vesicles in the low tension regime, a tension change $\Delta (\ln \sigma)$ imposed by a change of apparent area can be written as \cite{Rawicz2000}: 
\begin{equation}
\label{eqnlnsigma}
\ln\left( \Sigma/\Sigma_0\right) \approx (8\pi\kappa_b/k_BT)\times\Delta A/A. 
\end{equation}

Assuming $\Sigma_0 = 10^{-9}-10^{-10}$ N.m$^{-1}$, it would lead to  $\Delta A/A\approx 10^{-3}$, which is one order of magnitude lower than  the apparent area change $R_P^{2}/R_{v}^{2} \approx 10^{-2}$ in our experimental system. Hence, the area consumed by the first 3 penetrating particles does not translate into a decrease of the external spherical area of the vesicle.  
Indeed, low tension vesicles always show some membrane area reservoirs, stored as internal structures as seen in fluorescence microscopy (App. \ref{structures}) presumably stabilised by (negative) spontaneous membrane curvature $m$ as described for bilayers exposed to asymetric sugar solutions \cite{Lipowsky2012}.  If one accounts for the existence of a spontaneous curvature $m$, the force needed to pull a tube $f_{in}$ discussed earlier for the plateau force reads $f_{in}= 2\pi\sqrt{2\kappa\left(\Sigma+2\kappa m^2\right)}+4\pi \kappa m$. We can then interpret $F_{tube}=f_{in}$ data in two regimes. The first one for the first three  particles where the measured $F_{tube}$ remains constant and where only the membrane area from the reservoirs is consumed, which corresponds to the dashed line in Fig. \ref{fig2}C for $ \left|m\right|= 2 \times 10^5 $ m$^{-1}$.  
The second regime starts instead when the membrane area stored in the reservoirs is not accessible anymore and the mechanical membrane tension $\Sigma$ starts to increase as expected from Eq. (\ref{eqnlnsigma}), see Fig. 2C.

Now we focus our attention on the quantitative interpretation of the force profile data considering the energy landscape models (App. \ref{models}) composed of bending, tension and adhesion contributions, in the limit $R_v\gg R_P$ \cite{Deserno2004}. In these models only the membrane area bound to the particle contributes to the energy landscape, given that the unbound membrane close to the particle can adopt zero energy shapes. We calculate the force contribution from the energy by taking the derivative of the energy with respect to displacement of the contact line along the particle $s = R_P \alpha$. The energies expressed with the parameters of our system are 
 $E_{b}=  4\pi\kappa_b \left(1+mR_P\right)(1-\cos\alpha)$, $E_{\sigma}=\sigma \pi R_P^2 (1-\cos\alpha)^2$ and $E_{w}=w2\pi R_P^2 (1-\cos \alpha)$. Hence, 
the force is  $F= -dE/\left(R_Pd\alpha\right)$. The modulii of the force contributions acting on the contact line between the bound and unbound membrane regions then take the form :

\begin{eqnarray}
F_b=\frac{\kappa_b  4\pi \sin \alpha}{R_P}+4\pi\kappa_b m,
\label{eq:one}
\\
F_\sigma= 2 \pi R_P \sigma  \sin \alpha(1-\cos \alpha)\;,
\label{eq:two}
\\
F_w=2\pi R_P w \sin \alpha\;
\label{eq:three}.
\end{eqnarray}

We can postulate that the maximum force recorded upon penetration can be found by summing of the contributions in Eq. (\ref{eq:one} - \ref{eq:three}) at their maximum (around $\pi/2$) (App. \ref{models}). Note that again, $\sigma$ in $F_{\sigma}$ is an effective tension inferred from $F_{tube}$ accounting for $\sigma=\Sigma+\tilde{\sigma}$.

\begin{figure}[t]
\includegraphics[width=1\linewidth]{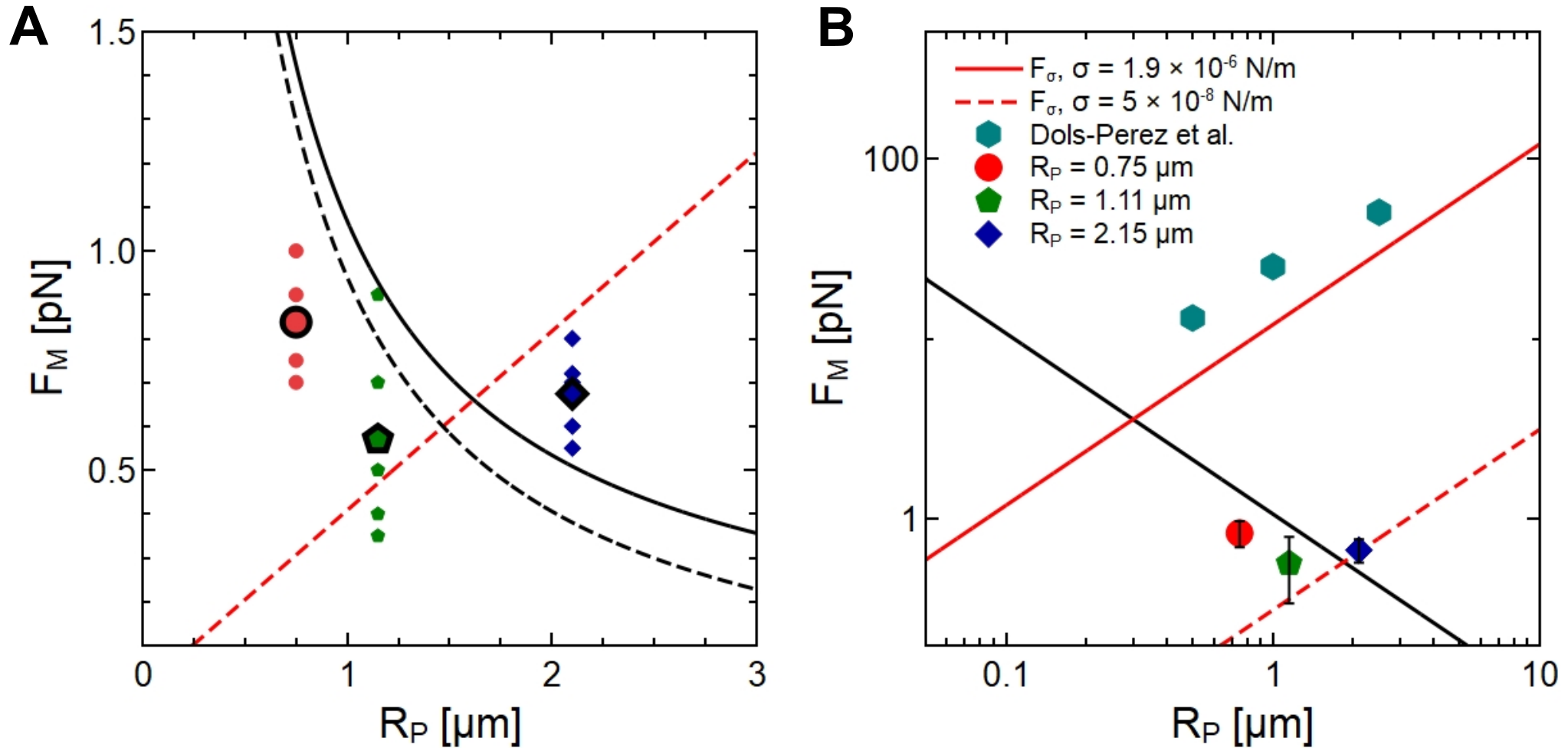}
\caption{\textit{(A)} Maximum force recorded upon penetration of Silica particles for different $R_P= 0.7, 1.1$ and $ 2.1 \ \mu$m. Black solid (dashed) curve is the maximum of $F_{b}(R_P)$ (Eq. (2)) for $m=0$ (and $m=-10^5  $ m$^{-1}$) . Red dashed curved is the maximum of $F_{\sigma}(R_P)$ (Eq. (3)). \textit{(B)} $F_M$ vs $R_P$ in log-log scale. Solid red line is the maximum of $F_{\sigma}(R_P)$ for $\sigma=1.9 \times 10^{-6}$ N.m$^{-1}$ measured in \cite{DolsPerez2019} with associated $F_M$. }
\label{fig3}
\end{figure}

\begin{figure}[b]
\includegraphics[width=1\linewidth]{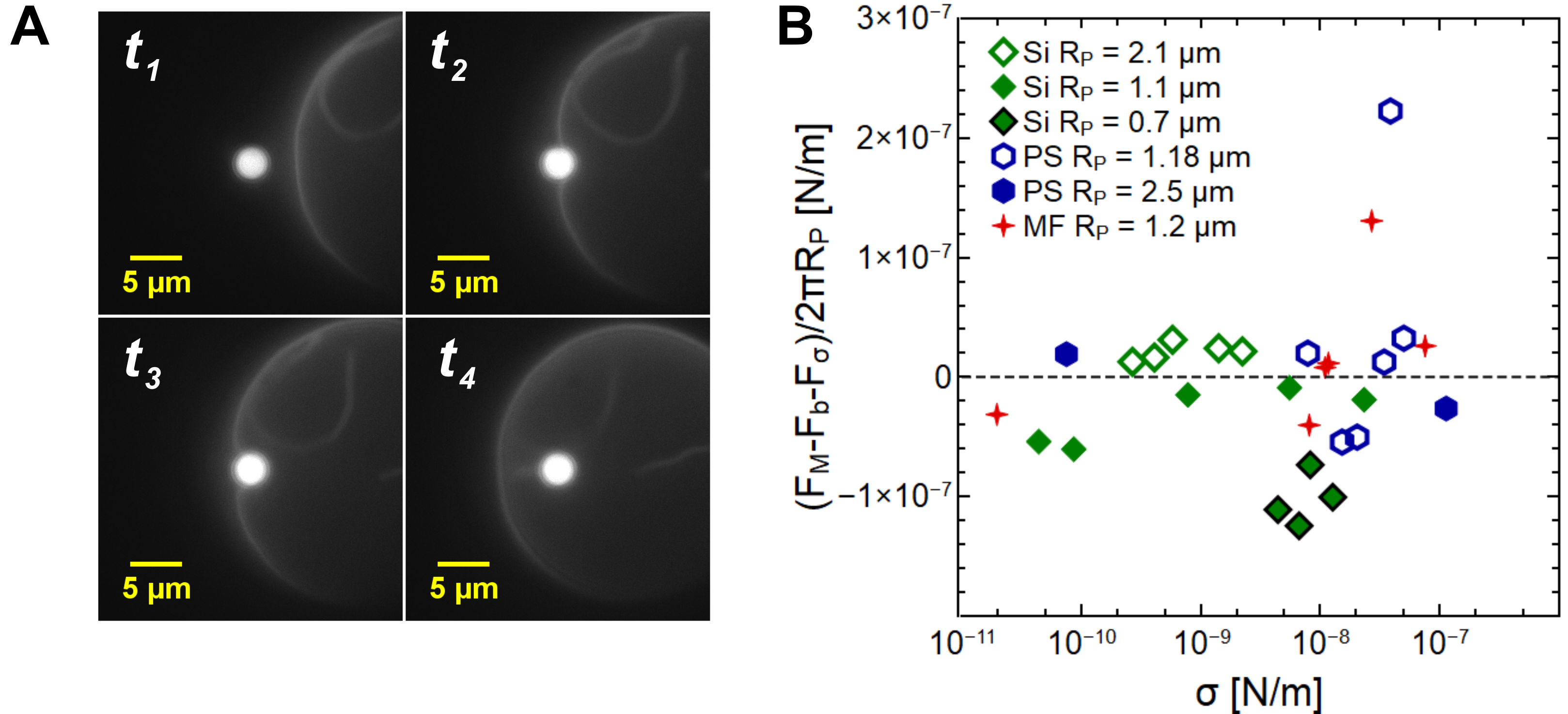}
\caption{\textit{(A)} Snapshots of a fluorescent $R_P=1.2 \ \mu$m Melamine Formaldehyde (MF) particle penetrating a GUV. \textit{(B)} Maximum force $F_M$ with subtracted contributions versus vesicle tension inferred from $F_{tube}$.}
\label{fig4}
\end{figure}

To investigate the bending contribution to the force profile, we  performed similar penetration experiments on different GUVs with a distribution of tensions around $\sigma = 10^{-8}$ N.m$^{-1}$ using three different Silica particle sizes $R_P=0.75 \ \mu$m, $R_P=1.11 \ \mu$m and $R_P=2.15 \ \mu$m. The results are shown in Fig 3A. As seen from Eq. (\ref{eq:one}), the bending contribution to the force acting on the penetrated particle is inversely proportional to the particle radius $R_P$. As the vertical dispersion of $F_{M}$ for individual measurements in Fig \ref{fig3}A stands for the dispersion in the membrane tensions of the penetrated GUVs, it appears that the averages as well as lowest measured values (where tension contributions are as small as possible) for the two smallest sizes follow a $\propto 1/R_{P} $ dependency. The fact that our measurements lay underneath the predicted maximum force needed to bend the membrane is a clear sign that there exists a weak adhesion energy facilitating the penetration of the Silica particle. For the largest particle radius $R_P=2.15 \ \mu$m however, the results do not seem to follow the trend if we only take into account the bending cost. Indeed, considering that the membrane area needed to wrap the particle scales with $\propto R_P^{2}$, the tension contribution ($F_\sigma \propto R_{P} $) when using large particles can no longer be disregarded. Hence, tension contributions ultimately play a role during the penetration of the particle resulting in higher $F_{M}$. In other situations for larger particles, the shape of the unbound region of the membrane may not correspond to a minimal surface as predicted in several theoretical models, which leads to additional force costs  \cite{Powers2002,Prost2002}.
In Fig \ref{fig3}B, we plot our results and the ones reported by Dols-Perez et al. \cite{DolsPerez2019}  operated in a tension-dominated regime, $R_P > \lambda_\sigma$ where $F_{M}$ scales linearly with $R_P$, $F_{M}\propto R_P$ (Eq. \ref{eq:two}) thereby evidencing the existence of the crossover regime. 
Indeed, as the bending-dominated regime (Eq. (\ref{eq:one})) is characterized by $F_{M}\propto R_P^{-1}$, a force barrier minimum can be observed in the crossover regime where the particle radius is comparable to the bendocapillary length. 

Finally, we performed experiments with particles made of different materials (Polystyrene, Melamine formaldehyde and Silica) that show different physico-chemical properties (e.g. Zeta Potential measurements App. \ref{particles}) and in turn different adhesion $w$ with the membrane \cite{Spanke2020,Sharma2022}. For low tension vesicles,  we were able to penetrate and stably wrap MF and PS particles using the same range of trapping forces as used for Silica particles and obtained similar force profiles (see Fig. \ref{fig4}A, Supplementary Video S4). Knowing that $F_{M}$ contains information about all contributions ($F_b+F_\sigma+F_w$), and being able to evaluate the membrane tension-related contribution $F_\sigma$ from $F_{tube}$, we can evaluate  $w$  to be $\left(F_{M}-F_b-F_\sigma \right)/2\pi R_P$, which is plotted as a function of $\sigma$ for the different particle types in Fig \ref{fig4}B. Note that $w$ remains weak ($<3 \times 10^{-7}$ N.m$^{-1}$) showing both signs, which corresponds to favourable or unfavourable adhesion, and that there is no clear correlation between the membrane tension and evaluated particle adhesion $w$. Hence, applying an external pN force leads to the membrane wrapping of even non adhesive particles. These results agree with the picture of a weak and non specific adhesion where  the interaction between the particle and the membrane is mediated by a water film of $h$ = 10-100 nm  thickness \cite{CardosoDosSantos2016}. Dispersion of the data shown in Fig. \ref{fig4}B could be interpreted in terms of a distance-dependent particle-membrane adhesion $w(h)$, which can result from the contributions of electrostatic double layer \cite{Ewins2022}, steric \cite{Spanke2020}, hydrophobic and Van der Waals interactions \cite{Bahrami2014}. 

In conclusion, we have performed force-driven penetration of spherical microparticles of various compositions into giant unilamellar vesicles by means of optical tweezers in the regime of low membrane tensions. We evidenced the negligible impact of the particle-membrane adhesion in this regime and showed that several particles can be penetrated in a same vesicle before triggering a significant tension-mediated resistance against particle wrapping. Doing so, we put forward the role of membrane excess area stored in the vesicle area reservoirs, stabilised by spontaneous membrane curvature, for the particle entry to occur. Finally, we show that a force barrier minimum exists for a particle size comparable to the bendocapillary length, whose magnitude depends only on the membrane properties. These experiments supported by models provide a precise quantification of the forces required for the internalization of particles implying morphological transitions of lipid membranes, which is relevant within the scope of drug vectorization applications and cellular endocytosis.

\begin{acknowledgments}
We wish to acknowledge  helpful  conversations and insightful comments from Carlos Marques as well as funding from the Ecole Doctorale Physique Chimie-Physique of Strasbourg, ANR EDEM (ANR-21-CE06-0042-01) and ITI HiFunMat (U. Strasbourg).
\end{acknowledgments}

\appendix

\section{Material and Methods}

\subsection{Lipids and Gel-assisted GUV formation method}
\label{lipids}

The Giant unilamellar vesicles (GUVs) used in this work were prepared using a PVA
(Polyvinyl alcohol) gel-assisted formation method.
The PVA gel is prepared by dissolving PVA in pure water (MilliQ water) at 5 \% concentration.
The prepared PVA gel is spread on a PTFE (Polytetrafluoroethylene) plate
and dried for 45 minutes at 80$^{\circ}$C in an oven. In the case of POPC/POPC-NBD vesicles,
5 $\mu$L of a 99:1 (molar) mixture of POPC (1-Palmitoyl-2-oleoylphosphatidylcholine) and
POPC-NBD (POPC fluorescently labelled with Nitrobenzoxadiazole) lipids in chloroform
(1 g/L) are spread on the PVA gel and vacuum dried in a desiccator for 15 minutes. At this stage, the lipids under solvant evaporation spontaneously form stacks of lipid layers supported by the dried PVA gel film.
This lipid system obtained is then hydrated with 200 $\mu$L of sucrose (150 mM) and allowed
to grow for 2-3 hours. The vesicle suspension is then collected and sedimented in 1
mL of glucose (150 mM) solution. The outer glucose concentration relative to the inner
sucrose concentration can be modifed to create a slight hypertonicity that will lead to
lower tension vesicle membranes. The slight density mismatch between the sucrose solution
inside the vesicle and the sucrose/glucose solution in the aqueous medium allows
the vesicles to sediment at the bottom of the cell without strongly deforming the vesicle.

\subsection{Particles}
\label{particles}

The particles used in this work are spherical non-porous Silica (SiO$_2$), Melamine formaldehyde (MF) and Polystyrene (PS) particles, all purchased  from microParticles GmbH. 

Zeta potential of the Silica and Melamine formaldehyde particles was measured using Malvern Zetasizer Nano ZS. The zeta potential of Silica particles was measured to be $\zeta_{Si} = -75 \pm 5$ mV while the one for MF reads $\zeta_{MF}=25 \pm 6$ mV. Measurements of Zeta potential of Polystyrene particles in water in the literature agree on $\zeta_{PS} =  -60$ mV to $\zeta_{PS} =  -40$ mV \cite{Zeta1,Zeta2,Zeta3}.

In order to prepare diluted particle dispersions to perform investigations, particles were transferred from the mother highly concentrated dispersion (5 \% (w/v) aqueous suspensions) to the sample cell by evaporating few microliters of the mother dispersion on a Silicon wafer. Using a micropipette previously filled with the corresponding medium of the experiment, particles were extracted from the particle coated Silicon wafer. Adding this volume to the sample cell, a diluted particle dispersion can be obtained for the experiments.

\subsection{Optical setup}
\label{microscopy}

Trapping laser source is a 976 nm single mode laser diode (Thorlabs CLD1015) with tunable output power up to 300 mW. The objective used is a high numerical aperture 100X Nikon Plan Fluorite Oil Immersion Objective, 1.3 NA, 0.16 mm WD. For both trapping the particles and imaging the sample a standard inverted microscope configuration taking advantage of a CMOS sensor camera Hamamatsu ORCA-Flash 4.0 C11440 was used.
Bright field illumination light source was a LED illumination system furnished by Thorlabs Inc.
The fluorescence microscopy module used is the Thorlabs OTKB-FL coupled with a Nikon C-HGFI Intensilight source.

\subsection{Sample cell preparation}
\label{sample}

The sample cell consists in a thin glass coverslip (0.17 mm thickness, Menzel-Gläser) on top of which a self-adhesive silicon imaging chamber (CoreWell Imaging Chamber) of 0.9 mm diameter and 1.6 mm thickness is placed. The whole forms a sealed through which can then be filled with 150 $\mu$L of a 0.15 Osm/kg glucose solution. 1-5 $\mu$L of concentrated GUVs solution can then be added as well as 5 $\mu$L of particles extracted from the particles-coated silicon wafer. The whole is then left open for an hour to allow evaporation of the water in the glucose phase and induce the deflation of the GUVs. 

\subsection{Influence of penetration speed}
\label{penetrationspeed}

\begin{figure}[ht!]
\includegraphics[width=0.6\linewidth]{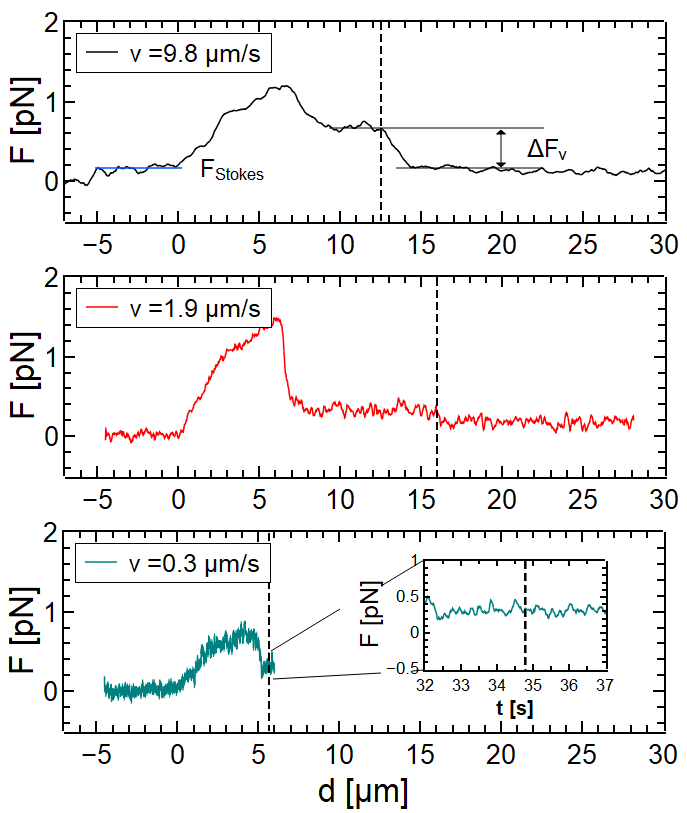}
\caption{Force profiles upon penetration of $R_P=1.15 \ \mu$m Silica particles performed at different velocities.Vertical dashed lines stand for the moment the sample stage stops its motion and velocity therefore drops to zero. Inset for the lowest velocity shows the tube pulling force at the moment the stage is stopped.}
\label{figvelo}
\end{figure}

The relative speed between the particle and the approaching vesicle could easily be tuned in our experiments using the interfaced piezoelectric actuators controlling the sample stage (Thorlabs 3-Axis NanoMax MAX381). In this work, we aim at probing the response and dynamics of the membrane/particle system and get rid of the effects arising from the approaching and penetration velocity. We performed penetration experiments at different velocities and plot the profiles to investigate the effect of the velocity in Fig \ref{figvelo}.

For high velocities, a force plateau is observed before contact accounting for the Stokes friction felt by the particle which scales linearly with velocity. This force therefore becomes negligibly small for lower velocities. The second striking influence of velocity arises in the plateau following the wrapping of the particle when the tube is being pulled. Indeed, at large velocities, a large force drop $\Delta F_v$ is measured between the tube pulling phase ($v \ne 0$) and the tube holding phase ($v = 0$, when the sample stage stops moving and vesicle as well as particle are at rest) showing the contribution of velocity when pulling the tube. Again, this $\Delta F_v$ becomes negligibly small for lower velocities leading to $F_{tube_{speed}}=F_{tube_{rest}}=f_{in}$. However, the force profile during the contact and deformation remains consistently similar for vesicles with comparable tensions in every velocity regime. For these reasons, all systematic experiments performed in this work were performed at $v = 0.3 \ \mu$m.s$^{-1}$.

\subsection{Irreversibility}
\label{irreversibility}

\begin{figure}[ht!]
\includegraphics[width=1\linewidth]{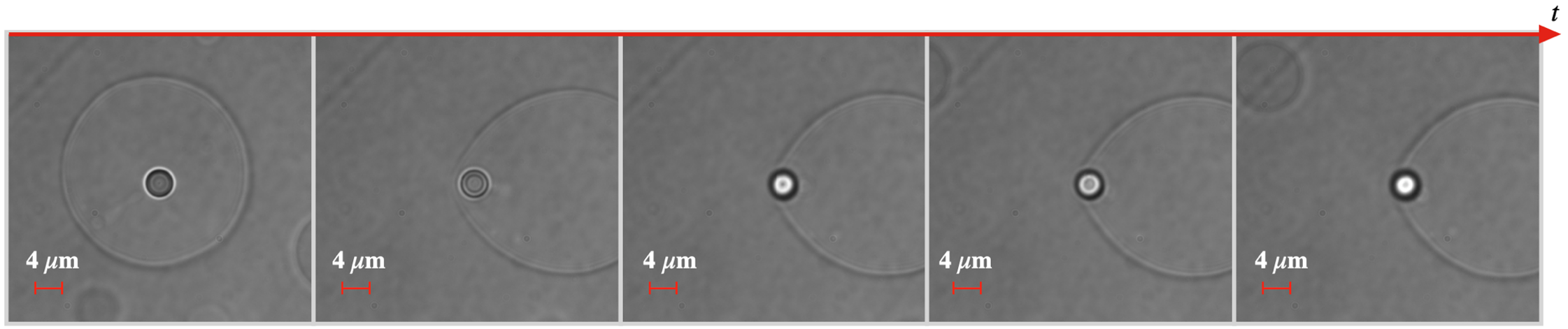}
\caption{Transportation of a GUV observed when trying to force the particle out after the particle was already penetrated inside with the optical tweezers.}
\label{figirrev1}
\end{figure}

After penetration, the particle is fully wrapped by the membrane and connected to the mother vesicle by a tube. In this configuration, we observe that it is impossible to perform the reverse process namely forcing the particle (for a same optical power) to unwrap and cross the membrane from the inside to the outside. All the attempts to perform this manipulation either led to the particle escaping the optical trap or to transportation of the whole GUV as shown in Fig \ref{figirrev1}.

When higher optical power was used (200 mW), we could observe the formation of an outward tube but the latter would always retract when the optical trap was turned off, bringing the particle back inside the GUV in the original state.

\subsection{Trap calibration and force measurements}
\label{calibration}

Precise calibration of the optical trap was performed using Boltzmann statistics on the trajectories of the trapped particle center of mass recorded with the camera. The tracking of the center of mass was achieved using the open-source software Blender using a Kanade-Lucas-Tomasi feature tracking algorithm. For long enough trajectories one can reconstruct the effective potential felt by the particle which is quadratic as expected for a particle submitted to the restoring force of an optical trap. The fitting of the parabola allows to read directly the trap stiffness $\kappa$ along a given dimension. 

\begin{figure}[ht!]
\includegraphics[width=0.6\linewidth]{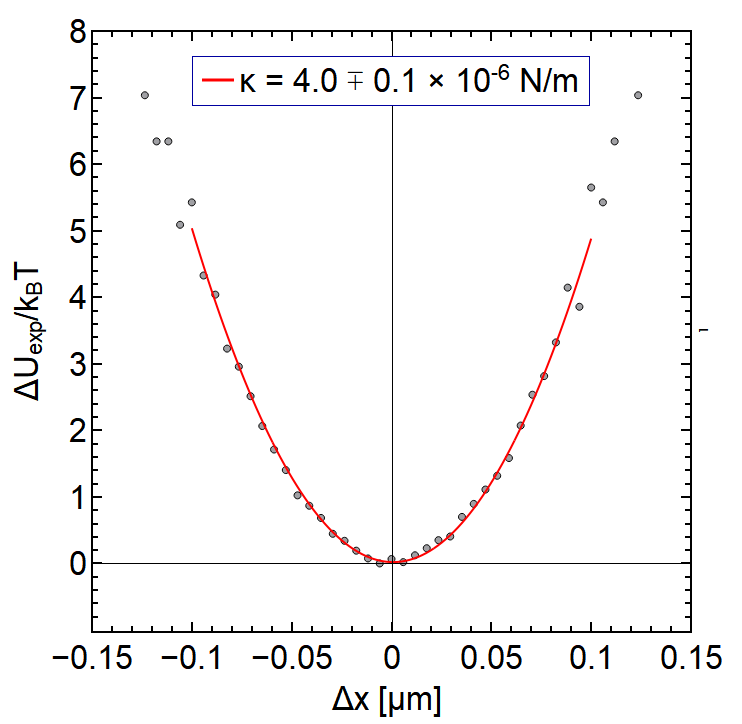}
\caption{Experimental effective potential along a single dimension felt by a $R_P=2.15 \ \mu$m Silica particle in an optical trap at 100 mA. Trajectory for $N$ = 15000 points.}
\label{figcal}
\end{figure}

Error on the determination of the trap stiffness $\kappa$ is evaluated from the fit mean squared error. A precision of $1 \times 10^{-7}$ N.m$^{-1}$ is achievable for calibration performed on trajectories with 15000 points or more.

\subsection{Modelling the force profiles}
\label{models}

\begin{figure}
\includegraphics[width=0.65\linewidth]{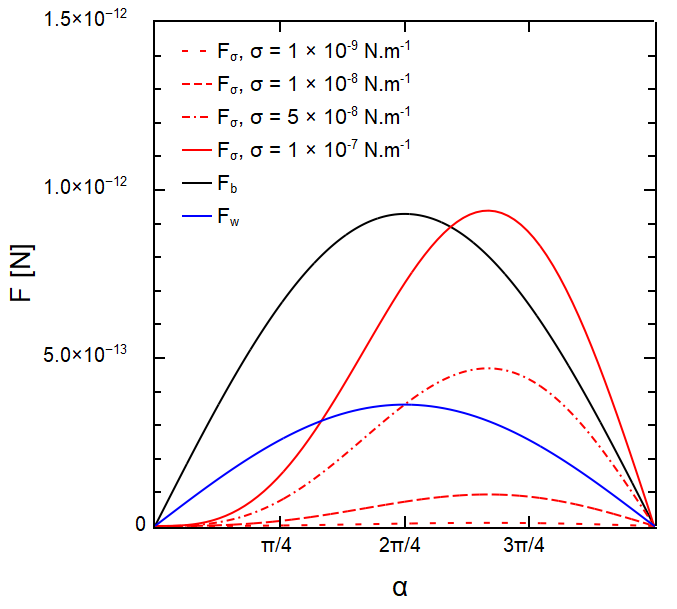}
\caption{Theoretical force profiles for each contribution of the model. $F_b$ is plotted with $\kappa_b =0.85 \times 10^{-19} $ J and $F_w$ with $w=5 \times 10^{-8}$ N.m$^{-1}$ (Eq. 2-4). Note that $F_w$ is positive here but can be positive or negative depending on the sign of $w$.}
\label{figmodel}
\end{figure}

The energy considerations derived in previous works \cite{Deserno2002} allow to have insights on the force that has to be provided for wrapping of the particle to be observed when taking the derivative with respect to the contact line displacement. In Fig. \ref{figmodel}, we plot the modulii of the forces as a function of contact angle (Eq. \ref{eq:one}-\ref{eq:three}). 
It is worth noting that for the bending and adhesion contributions, the extrema are at $\alpha=\pi/2$ which is not the case for $F_{\sigma}$. The maximum happens at $\alpha_{max}$ which is such that $\sin \alpha(1-\cos\alpha)$ is maximum giving $\alpha_{max}=3\sqrt{3}/4$.

It is worth noting that throughout the analysis, only the modulii of the forces are considered. However in general, the force vectors for the tension or bending contributions are not collinear with the optical trapping restoring force. The considered contributions should then be the projections of the force vectors on the axis of axisymmetry of the problem. Still, due to the difficulty to accurately define a contact angle with bright field images together with the fact that the maximum of the contributions happen to occur around $\alpha = \pi/2$ where the vectors are collinear with the optical trap restoring force,we use the modulii of the forces to calculate the contributions. 

\subsection{Internal structures}
\label{structures}

\begin{figure}
\includegraphics[width=0.85\linewidth]{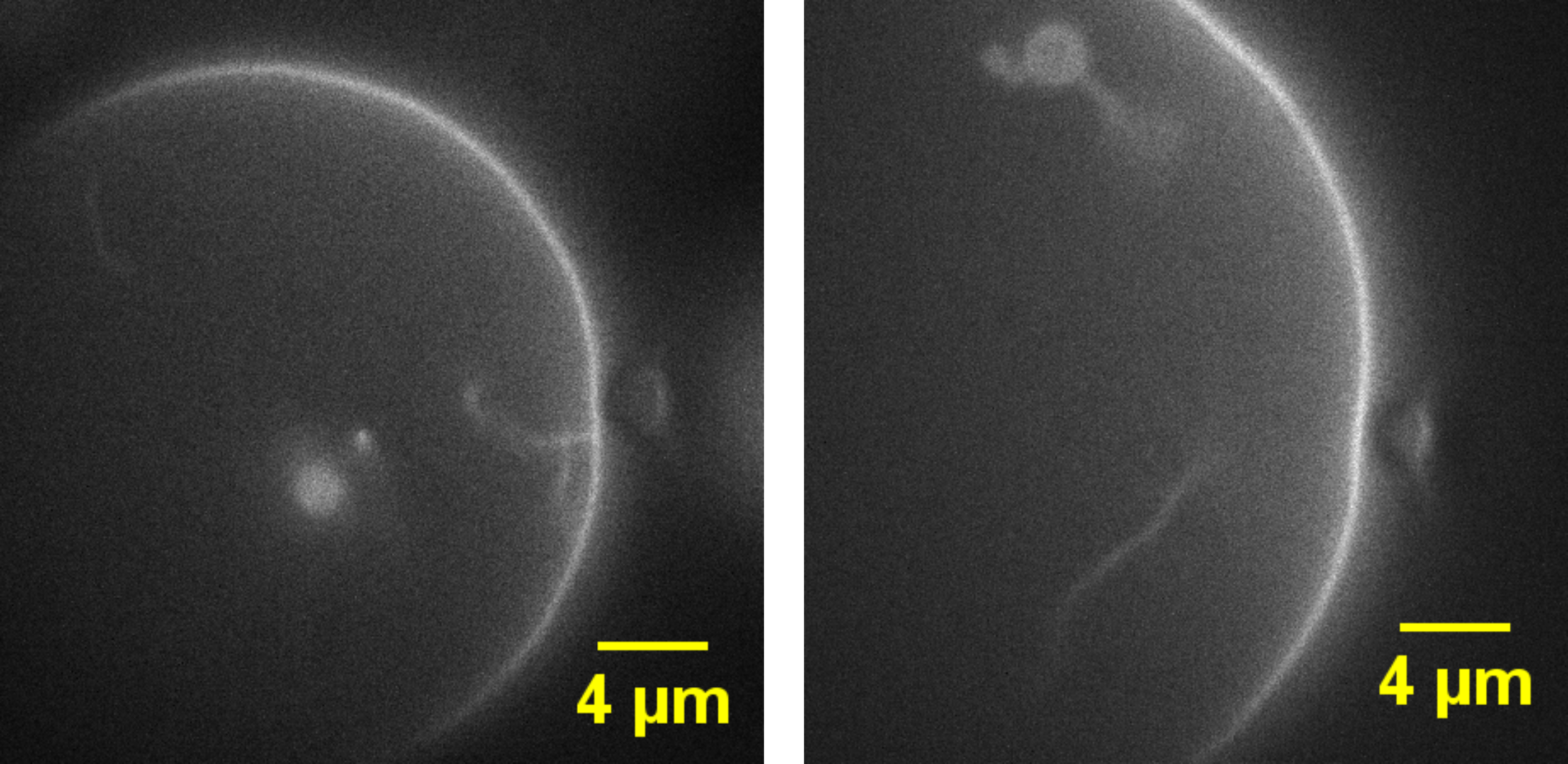}
\caption{Fluorescence microscopy snapshots of POPC GUVs showing internal structures after osmotic deflation, before a particle penetrates.}
\label{figstruct}
\end{figure}

After osmotic deflation of the GUVs by exposing them to lower concentration surrounding glucose medium (upon  evaporation), the vesicles start to show apparent floppiness giving evidence of a decrease in membrane tension. In addition to that, fluorescence microscopy allowed to acknowledge the existence of internal structures such as tubes or inner buds as shown in Fig \ref{figstruct}.

The existence and stability of these structures point towards the fact that a spontaneous curvature might exist in our system \cite{Lipowsky2012}. In addition, the fact that all of them are internal and almost no external structure was observed suggests that the sign of this curvature is negative.

\section{Supplemental Material}

\subsection{Videos}
\label{videos}

\begin{itemize}
    \item Video S1 : Fluorescent microscopy recording of an optically trapped $R_P=1.15 \ \mu$m Silica particle penetrating a NBD-labelled POPC GUV. 
    \item Video S2 : Bright field microscopy recording of an optically trapped $R_P=1.15 \ \mu$m Silica particle penetrating a POPC GUV. 
    \item Video S3 : Bright field microscopy recording of an optically trapped $R_P=1.15 \ \mu$m Silica particle penetrating the same POPC GUV as in Video S2 after three other particles were penetrated before.
    \item Video S4 : Fluorescent microscopy recording of an optically trapped $R_P=1.17 \ \mu$m PS particle penetrating a NBD-labelled POPC GUV.
\end{itemize}

\bibliography{apssamp}

\end{document}